\documentclass[
    ,final            
  ]
 {aipproc}

\layoutstyle{8x11single}

\def\be{\begin{equation}}
\def\ee{\end{equation}}

\def\lsim{\lower 2pt \hbox{$\, \buildrel {\scriptstyle <}\over
         {\scriptstyle \sim}\,$}}
\newcommand\gsim{\buildrel > \over \sim}

\begin{document}

\title{Pulsar twinkling and relativity}


\author{Isabelle A. Grenier}{
  address={AIM, Service d'Astrophysique, CEA Saclay, 91191 Gif/Yvette, France}
}

\author{Alice K. Harding}{
  address={NASA Goddard Space Flight Center, Greenbelt, MD 20771, USA}
}

\begin{abstract}
The number of pulsars with detected emission at X-ray and
$\gamma$-ray energies has been steadily growing, showing that
beams of high-energy particles are commonly accelerated in pulsar
magnetospheres, even though the location and number of
acceleration sites remain unsettled. Acceleration near the
magnetic poles, close to the polar cap surface or to higher
altitudes in the slot gap along the last open field lines,
involves an electric field component due to inertial-frame
dragging. Acceleration can also take place in the outer
magnetosphere where charge depletion due to global currents causes
a large electric field along the magnetic field lines. All models
require a detailed knowledge of the open magnetosphere geometry
and its relativistic distortions. Observational trends with age,
spin-down power and magnetic field as well as population synthesis
studies in the Galactic disc and the nearby Gould Belt provide
useful, however not yet conclusive, constraints on the competing
models.
\end{abstract}

\maketitle

\section{Introduction}
Pulsar twinkling is all relative. It depends on the shape of the
pulsar magnetic field and the observer's point of view, on the
metric near the neutron star and on retarded potentials and
aberration of light in the outer magnetosphere. General and
special relativity are involved at three stages: (1) for the
unipolar inductor to extract charges and draw currents above the
polar caps in dragged inertial frames; (2) to distort the magnetic
field, thus affecting the pair cascading efficiency and the
high-energy curvature radiation spawned by the primary charges;
(3) to control the location and extent of the accelerating sites
(gaps), thus lightcurve morphologies and polarization patterns.
The overall flux and spectral distributions predicted as input to
population synthesis studies are therefore quite sensitive to
these effects.

Over 1560 radio pulsars have been discovered with periods ranging
from milliseconds to seconds according to their history, and with
spin-down powers spanning 10 orders of magnitude, from $10^{21}$
to $10^{31}$ W. The bulk of the population is middle-aged, around
$10^{6-7}$ years, with canonical fields near $10^8$ T.
Billion-year old millisecond pulsars have weak $10^{4-5}$ T
fields. The slow $10^{4-6}$ year-old magnetars have supercritical
fields of $10^{10-11}$ T. Pulsars seen at high energy often have
large fields in their outer magnetosphere, either because of their
intense stellar field, as in anomalous X-ray pulsars, or because
of the compactness of their magnetosphere, as in ms pulsars, or
their youth for $\gamma$-ray pulsars. The latter are ostensibly
younger than several $10^5$ years because of the limited
sensitivity of the current $\gamma$-ray telescopes. Despite the
large size of the radio sample, little is known about the coherent
process responsible for the radio pulses which consist of a core
component near the polar cap and a hollow cone component which
probably takes place at several tens of stellar radii above the
polar cap ($r_{radio}/R \sim 40 \nu_{GHz}^{-0.26} P^{0.3}$ at 
frequency $\nu_{GHz}$ in GHz, for a
neutron star radius R and period P in seconds \cite{kijak03}).
More straightforward, incoherent processes, like curvature and
synchrotron radiation and Compton scattering, give rise to the
optical, X-ray and gamma-ray pulses.

The observed pulsed emission depends on a small set of parameters:
the pulsar angular velocity \textbf{$\Omega$}, its spin-down power
$\dot E_{\rm psr} = I \Omega \dot{\Omega}$ for a moment of inertia $I$, its
characteristic age $\tau = \Omega /(2\dot{\Omega})$; the
inclination angle $\alpha$ of its magnetic field to the rotation
axis and the observer's viewing angle $\zeta$ to the same axis;
the radius of the light cylinder $R_{LC} = c/\Omega$ where the
corotation velocity reaches the speed of light; the intensity of
the surface magnetic field $B_p$ at the pole; the polar cap half
angle $\sin \theta_{PC} = (R \Omega /c)^{1/2}$. The Deutsch vacuum
solution\cite{deutsch55} for the field is often adopted.
The near-dipole geometry
has been recently confirmed by \cite{lyutikov05}.

\section{Electrodynamics and acceleration}
Rotating, magnetized neutron stars are natural unipolar inductors,
generating huge electric fields in vacuum ($\vec{E} = -(\vec
\Omega \wedge \vec{r}) \wedge \vec{B}$) and building a large
surface charge. However, Goldreich \& Julian \cite{goldreich69}
noted that the electric field component parallel to the magnetic field,
$E_{\parallel}$, at the stellar surface can pull charges out of
the star against gravity to build a force-free density in the
magnetosphere. Wherever the charge can reach the Goldreich-Julian
(GJ) charge density, $\rho_{GJ} \simeq - 2\epsilon_0 \vec{\Omega}
\cdot \vec{B}$, it is able to short out $E_{\parallel}$, and both
charges and dipole field will corotate with the star. Corotation
must break down at large distances from the neutron star due to
particle inertia. Global magnetospheric simulations \cite{petri02,
spitkovsky06} have not yet been able to show whether and how a
pulsar magnetosphere reaches the nearly force-free (ideal MHD)
state envisioned by Goldreich \& Julian, in particular how the
large amount of charges are supplied to meet the force free
conditions. Clearly, particle acceleration requires at least some
local departure from force-free conditions. So, a real pulsar
magnetosphere must exist somewhere between the idealized vacuum
and force-free states we have been able to study so far on a
global scale. Pulsar particle acceleration has so far been studied
on a local scale and two types of accelerator, polar cap and outer
gap, have received the most attention.

\subsection{Polar cap and slot gap accelerators}

In polar cap accelerators, voltage develops along open field lines
near and above the polar cap surface. The two main subclasses are
vacuum gap models \cite{ruderman75}, where charges are trapped in
the neutron star surface layers by binding forces and a region of
vacuum forms above the surface, and space-charge limited flow
(SCLF) models \cite{arons79}, where charges are freely emitted
from the surface layers and a voltage develops due to the small
charge deficit between the real charge density $\rho$ and $\rho
_{GJ}$ according to $\vec{\nabla} \cdot \vec{E_{\parallel}} =
(\rho - \rho _{GJ})/\epsilon_0$. The two types of accelerators
differ only by the surface boundary condition on the charge
density, where $\rho (R) = 0$ for the vacuum gap, and $\rho (R) =
\rho _{GJ}$ for SCLF accelerators. For vacuum gaps, $E_{\parallel}
= \Omega B_p R$ is the full vacuum value at the surface.  For SCLF
accelerators, $E_{\parallel} = 0$ at the surface but it grows with
increasing $r$ because $\rho$, which must satisfy charge
continuity along each field line, decreases as $r^{-3}$ while
$\rho _{GJ}$ decreases more slowly. The form of $E_{\parallel}$ in
SCLF accelerators is thus sensitive to the detailed distribution
of the charge density, which depends both on the open field line
geometry as well as the compactness of the neutron star.  At
altitudes $\{z \ll \theta_{PC}$, $z \gg \theta_{PC}\}$, with $z
\equiv (r/R - 1)$ being the height above the surface,  \be
\label{Epar} E_{_{||} } \simeq B_p \theta_{_{PC}}^2 \left[\{{z,
\theta _{_{PC}}^2 \left({r \over R}\right)^{-4}}\} \kappa \cos
\alpha + \{z, \theta _{_{PC}}^2 \left({r \over R}\right)^{-1/2}\}
\frac{\theta _{_{PC}}}{2} \sin \alpha \cos \varphi \right][1 -
\left({\theta \over \theta_{_{PC}}}\right)^2] \ee
\cite{muslimov92, harding98}, where $\theta$ and $\varphi$ are the
magnetic polar and azimuth angles, $\kappa = 2 G I /(c^2 R^3)$ is
the stellar compactness parameter, and $I$ the neutron star moment
of inertia. The first term in Eqn (\ref{Epar}) is due to inertial
frame dragging near the neutron star surface, and dominates for
small $r$ and low inclination, while the second term is due to the
flaring of the field lines.

The potential drop available for particle acceleration is limited
by the development of electron-positron pair cascades which screen
the $E_{\parallel}$. In vacuum gap models, the pair cascade is
initiated when the gap height becomes comparable to the photon
mean free path for one-photon pair creation in the strong magnetic
fields, and causes a sudden discharge of the vacuum. The potential
drop in the gap thus oscillates between $V_{vg} \sim \Omega B_p
(R\theta_{_{PC}})^2/2$ and 0. In SCLF models, the pair cascades do
not cause a discharge, but develop only at the upper boundary of
the accelerator, screening the $E_{\parallel}$ in a relatively
thin region above a pair formation front (PFF) by trapping a small
fraction of positrons that accelerate downward. These accelerators
can thus maintain a steady current of upwardly accelerating
electrons, at $j^{-}_{\parallel} \simeq c\rho _{GJ}$, and a downward
current of positrons, at $j^{+}_{\parallel} \ll c\rho _{GJ}$,
which heat the polar cap.  The accelerator voltage is determined
by the height of the PFF, which is again roughly comparable to the
pair creation mean-free path.

The geometry of the polar cap accelerator is determined by
$E_{\parallel}$ and the physics of pair screening. Near the
magnetic pole, $E_{\parallel}$ is relatively strong and the PFF is
very near the neutron star surface.  But at the polar cap rim,
which is assumed to be a perfectly conducting boundary,
$E_{\parallel}$ vanishes.  Near this boundary, the electric field
is decreasing and a larger distance is required for the electrons
to accelerate to the Lorentz factor needed to radiate photons
energetic enough to produce pairs.  The PFF thus moves up and
curves upward as the boundary is approached, forming a narrow slot
gap near the last open field line \cite{arons83}. Since
$E_{\parallel}$ is unscreened in the slot gap, particles continue
to accelerate and radiate to high altitude along the last open
field lines. High-energy radiation can therefore come from
low-altitude ($r < 2R$) pair cascades near the pole
\cite{daugherty96}, higher-altitude ($r < (3-4)R$) pair cascades
on the inner edge of the slot gap \cite{muslimov03}, and high
altitude ($r \sim (0.1 - 0.8) R_{LC}$) radiation from primary
particles in the extended slot gap \cite{muslimov04}.

\subsection{Outer gap accelerators}

Outer gap accelerator models \cite{cheng86,romani96} focus on
regions in the outer magnetosphere that cannot fill with charge,
since they lie along open field lines crossing the null surface,
$\vec \Omega \cdot \vec B = 0$, where $\rho _{GJ}$ reverses sign.
Charges pulled from the polar cap cannot populate the region
between the null surface and the light cylinder, and a vacuum gap
forms.  This assumes of course that the charges coming from the
polar cap on these field lines  do not undergo enough pair
cascading to screen the outer gap. If outer gaps form, they can
accelerate particles to high energy and the radiated $\gamma$-rays
can produce pairs by interacting with thermal X-rays from the
neutron star surface. The density of such X-ray photons is very
small in the outer gaps, but is enough to initiate pair cascades
since the newborn pairs accelerate in the gap, radiate, and
produce more pairs.  The gap size is limited by the pair cascades,
which screen the gap electric field both along and across field
lines, thus determining the emission geometry. Young pulsars,
having hotter polar caps and higher vacuum electric fields, tend
to have narrow gaps stretching from near the null surface to near
the light cylinder \cite{cheng94} while the gaps of older pulsars,
having lower electric fields, are much thicker and grow with age
\cite{zhang97}. When the gap fills the whole outer magnetosphere
(at $\tau \gsim 10^7$ yr) it ceases to operate, so that not all
radio pulsars can emit $\gamma$ rays. Death lines in $P$-$\dot P$
space predict which pulsars can sustain outer gaps, depending on
whether the X-ray photon field comes from cooling of the whole
stellar surface or from polar caps heated by the energy deposited
by the return flux of charges \cite{zhang04}.

Recent outer gap models \cite{takata06} solve Poisson's Equation
in two dimensions (along and across the magnetic field). Such 2D
models derive a somewhat different gap geometry than the classic
one-dimensional models. Depending on the amount of current flow
into the gap, of order 10-20 \% of the GJ current, the gap can be
long and narrow or wide and thick. Such current flow is required
for the gap to produce a sufficient high-energy luminosity
\cite{hhs03}. The 2D gaps can extend below the null charge surface
and to a maximum height of about $0.8 R_{LC}$, in contrast to the
classic outer gaps that were assumed to extend from the null
surface to $R_{LC}$.

\begin{figure}[t]
\vspace{2.0cm}
{\includegraphics[width=0.8\textwidth]{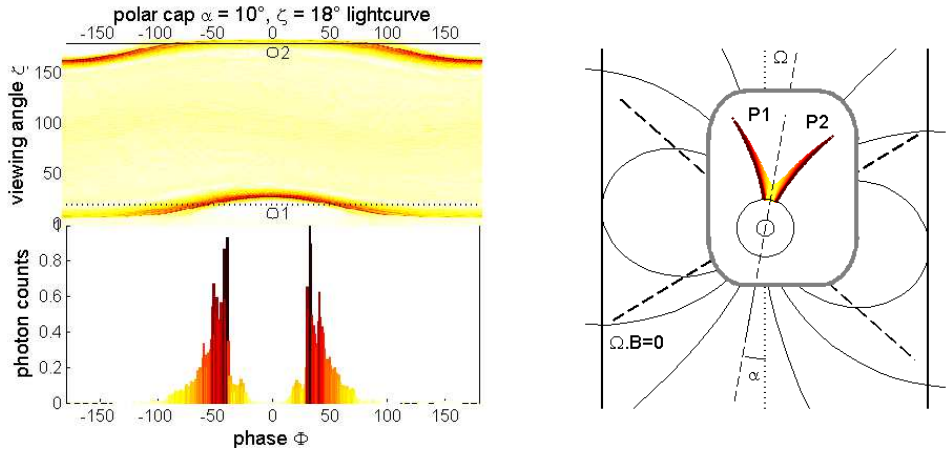}
\caption{Phase plot, sample lightcurve, and a sketch of the
accelerator location for the polar cap model, for a typical
inclination angle $\alpha = 10^\circ$. The central zoom gives the
gap extent relative to the star size. The dashed lines outline the
null surface. The shading in the lightcurve and gap sketch is the
same. The phase plot illustrates the change in lightcurve as seen
by different observers and the aperture of the pulsed beams.}
\label{figPC}}
\end{figure}

\section{Cascades and radiation}

The polar cap/slot gap and the outer gap models both use curvature
radiation to produce the primary $\gamma$ rays with energies
around 50 GeV to initiate the cascades \cite{hirotani01}
\cite{daugherty96}, but they differ in the pair creation process:
magnetic one-photon production in the intense field at low
altitude, and two-photon creation in the outer regions. Due to
screening of the $E_{\parallel}$ by the pair cascades, the maximum
Lorentz factor $\gamma_{e,max}$ saturates at several $10^7$ for
both the polar cap and outer gap and is very insensitive to
$\Omega$ and $\dot{\Omega}$. The following stages of the
photon-e$^{\pm}$ shower use the same radiation processes, namely
synchrotron radiation which is more efficient in the inner
regions, and inverse Compton scattering of the stellar thermal
radiation.

Polar cap pair cascades are initiated by primary curvature
radiation (CR), in the case of the young pulsars ($\tau \lsim
10^7$ yr), and by inverse-Compton scattering (ICS) of stellar
thermal X-rays by primary electrons \cite{sdm95}, for older
pulsars that cannot produce pairs from CR because of the weaker
$E_{\parallel}$ and straighter field lines \cite{hm02}. Secondary
to primary number ratios from CR cascades reach $10^3 - 10^4$
\cite{daugherty96}, but only $1-100$ for ICS cascades \cite{ha01}.
Radio emission is predicted to cease when a pulsar can no longer
even produce pairs from ICS \cite{hmz02}. The polar cap cascade
spectra of young, high-field pulsars show super-exponential
high-energy cutoffs at energies between 20 MeV and 20 GeV due to
magnetic pair attenuation. The cutoff energy is also influenced by
GR effects such as light bending and photon red-shifts
\cite{gh94}. The cutoff energy is expected to be lower on the
leading edge of the pulse due to rotationally induced asymmetries
in the pair absorption (larger angles between the photons and
field lines on the leading side) \cite{dr02}. The predicted
spectra of millisecond pulsars with very low surface fields are
not attenuated and may extend as high as 50 GeV \cite{hum05}.

Outer gap cascades are initiated by CR from primaries and can be
as rich as the polar cap cascades. Flux and spectra are quite
sensitive to attenuation by pair production and to the feedback
between the heated polar caps and the cascade development
\cite{takata06}. A distinctive feature of the outer gap is the
significant 0.1-10 TeV emission component due to Compton
scattering of soft IR and X-rays by the gap accelerated particles.
The outer gap and slot gap spectra show simple exponential
cut-offs due to the radiation-reaction limit of the particle
energy.  The return particle current appearing near the edge of
the open volume in global MHD simulations \cite{spitkovsky06}
could have some effect of the structure of both outer gaps and
slot gaps.

\begin{figure}[tbh]
\vspace{-2.0cm}
{\includegraphics[width=0.52\textwidth]{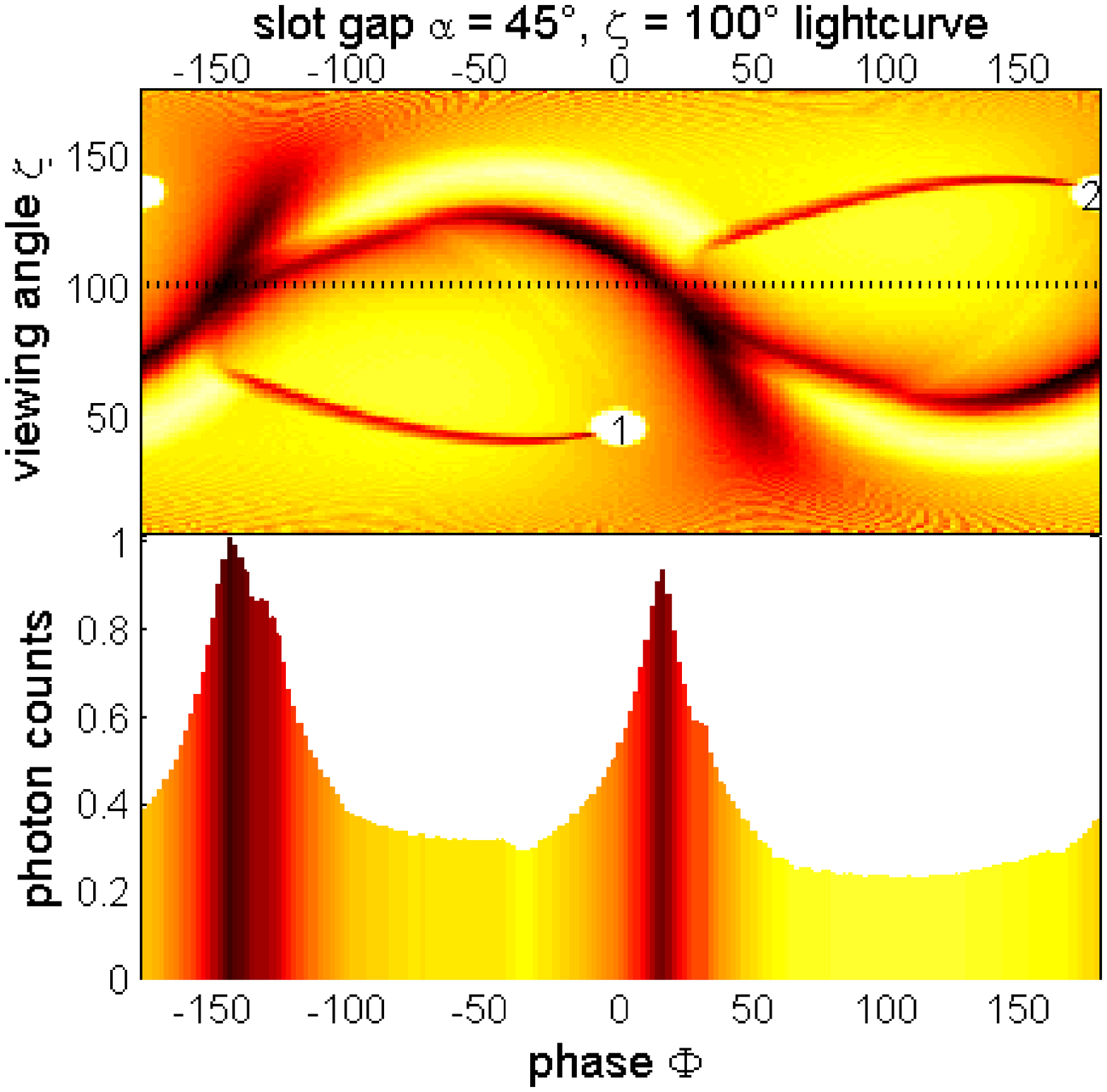}}
\vspace{-5.0cm}
{\includegraphics[width=0.32\textwidth]{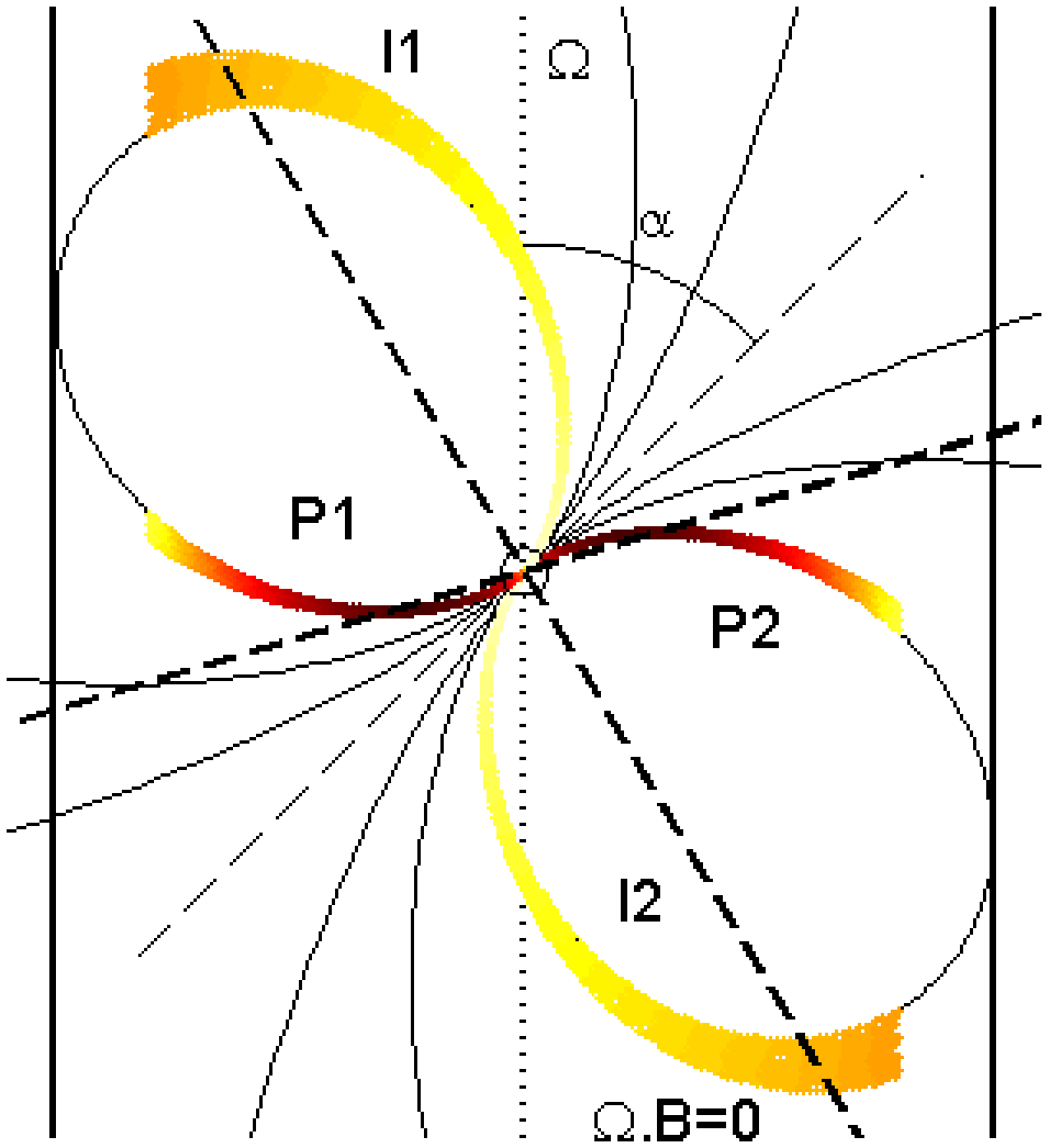}
\caption{Same as Figure \ref{figPC} for the slot gap model, for a
typical inclination angle $\alpha = 45^\circ$.} \label{figSG}}
\end{figure}

\section{Geometry and phase}
Assuming outward radiation tangent to the field lines, lightcurve
morphologies depend on general relativity (field distortion and
light bending near the star surface) and on special relativity
(light aberration, time-of-flight delays and field retardation
near the light cylinder). Aberration and time-of-flight produce
phase shifts of comparable magnitudes $\Delta \Phi \sim -r/R_{LC}$
in radiation emitted at different altitudes. On the leading side,
these phase shifts add up to spread photons emitted at various
altitudes over $\Delta \Phi \sim 0.4$ in phase. On the trailing
side, photons emitted later at higher altitudes catch up with
those emitted earlier at lower altitude. They arrive at an
observer within a small phase range $\Delta \Phi \sim 0.1$ and
produce caustics in the phase plot and light curve \cite{dr03, crz00}. 
These effects
dominate over the small degree of sweep-back of the retarded field
lines near the light cylinder ($\Delta \Phi \sim 1.2 (r/R_{LC})^3
sin^2 \alpha)$ \cite{dh04}.  The gravitational bending of light
paths near the surface is small and compensated by the reduced
size of the polar cap in a Schwarzschild metric
($\theta_{PC}^{GR}=\theta_{PC}(1+2GM/Rc^2)^{-1/2})$ \cite{gh94}.
The shape of the open volume depends on the pulsar obliquity and
retarded field (the polar cap radius varies with azimuth around
the pole), the pulsar age (the polar cap shrinks with age as
$\sin(\theta_{PC}) \propto \Omega^{1/2}$ and the light cylinder
expands as $R_{LC} \propto \Omega^{-1}$), and to a lesser degree,
on the reduced polar cap in a curved space-time. Currents can
easily further distort the weak magnetic field in the outer
magnetosphere, but also near the star. The lines swell along the
rotation axis under the drift ($\vec E \wedge \vec B$) current and
they curl more than the rotational sweep-back because of the
returning polar current (if electrons are extracted)
\cite{muslimov05}. All these effects strongly affect the width and
symmetry about the pole of the polar cap beam as well as the slot
and outer gap lengths and their peak phases for radiation.

\begin{figure}[tb]
{\includegraphics[width=0.8\textwidth]{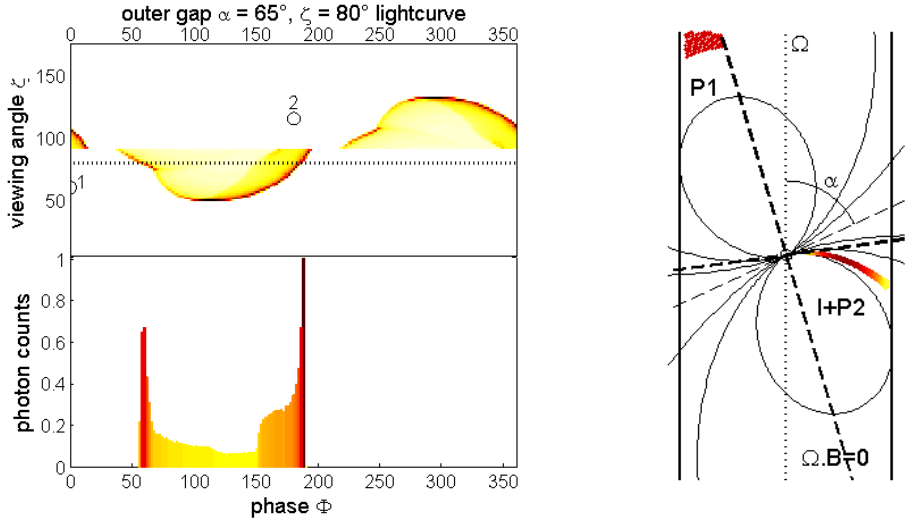}
\caption{Same as Figure \ref{figPC} for the outer gap model, for a
typical inclination angle $\alpha = 65^\circ$. The outer gap lies
along field lines with $\theta = 0.85 \theta_{PC}$ as in
\cite{crz00}.} \label{figOG}}
\end{figure}

Figures \ref{figPC} to \ref{figOG} try to capture these effects.
They show the dependence of photon intensity on phase for
different observer viewing angles, for the three models with
typical magnetic inclinations. The dipole sketches (simple,
unretarded dipoles) qualitatively illustrate the gaps location and
extent. The grey shading outlines the emerging photon phase in the
lightcurves as well as across the gaps.

A single polar-cap beam can produce a variety of pulse profiles
with any peak separation between 0 and 180$^{\circ}$ as long as
$\alpha \sim \zeta \leq 30^{\circ}$  (Fig. \ref{figPC}). Because
$E_{\parallel}$ fades away near the perfectly conducting edge of
the open volume, the gap is shorter near the pole ($0.5 R_*$) and
extends to higher altitude near the rim. Cascade synchrotron
emission is brighter along the more curved lines near the rim.
Faint and soft "off-beam" curvature radiation from the primary
particles above the gap can be seen outside the main beam, in
particular at large viewing angles \cite{hardingzhang01}. Off-beam
emission is softer because it originates inside the open volume at
large altitudes along field lines with larger curvature radii, and
because the particles have lost much of their initial energy.

Slot gap emission fills the whole sky and all phases in a
lightcurve. Most observers will catch emission from the two poles
if $\alpha \geq 30^{\circ}$ (for $45^{\circ} \leq \zeta \leq
125^{\circ}$ on Figure \ref{figSG}). The dark features show the
accumulation of photons because of the trailing side caustics (for
instance, the thick black curve at $\zeta < 130^{\circ}$ and
$|\Phi| < 50^{\circ}$ behind pole 1), and because of the overlap
between the trailing side of pole 1 and the leading side of pole 2
near the light cylinder (for instance, the thinner branch of the Y
feature at $50^{\circ} < \zeta < 90^{\circ}$ and $50^{\circ} <
\Phi < 110^{\circ}$). The main peaks come from the trailing side
of each pole, interpeak emission from the leading sides. The thin
arcs of emission emanating from each pole are caused by notches in
the polar caps distorted by retardation \cite{dh04}, which
produces bunching of field lines. Most of the emission takes place
at $0.1 \leq r/R_{LC} \leq 0.8$.

Being a subset of the slot gap above the null surface, an observer
can see only one pole from the outer gap (see Figure \ref{figOG}).
There is no emission outside the sharp peak edges and over a large
fraction of the $\Phi -\zeta$ space. The gaps are invisible at
$\zeta < 30^{\circ}$ or $\zeta > 150^{\circ}$ for any obliquity
and they shine only near the equator for small $\alpha$
inclinations. The dome-like structure in each half of the phase
plot is due to the shell-like shape of the trailing field lines.
The leading side emission shows up as a smaller dome protruding at
lower phase and partially overlapping with the trailing side.
Photons emitted on trailing field lines bunch up to form the
second peak caustic (similar to the slot gap peaks) while the
first peak originates near $0.9 R_{LC}$. The latter is very
sensitive to the assumed gap geometry (length along the lines,
thickness across the lines, and height above the last closed
lines). It would also disappear for emission very near the edge of
the open volume ($\theta = 0.9 \theta_{PC}$). It is also very
sensitive to the current feedback on the true field configuration.
Recent calculations \cite{takata06} show that the outer gap can
extend below the null surface, so wider beams and 2-pole emission
are possible as for the slot gap, but the electrodynamics in the
slot and outer gap models are quite different.

The phase plots outline the different beaming fractions of the
three emission models and the increasing probability of observing
radio-quiet objects from the polar cap to the outer gap model
because of the small aperture of the radio conal beam near the
poles. In the slot and outer gaps, peaks result from the trailing
caustic, so one expects well synchronized peaks across the entire
spectrum. Phase shifts between the radio core component and the
centroid of the conal peaks, as well as polarization patterns,
have been used to estimate the altitude of the radio emission: the
faster the pulsar, the closer the radiation is to the light
cylinder \cite{kijak03}. So, radio waves born on the trailing side
high enough to be in the caustic zone would appear in phase with
the high-energy photons \cite{hardingtexas05}. Hard X-ray
polarization will be a crucial tool to test the caustic role as it
implies a drop of the degree of polarization and a double swing by
$180^{\circ}$ of the position angle within the peak \cite{dhr04}.

\begin{figure}[tbh]
{\includegraphics[width=0.5\textwidth]{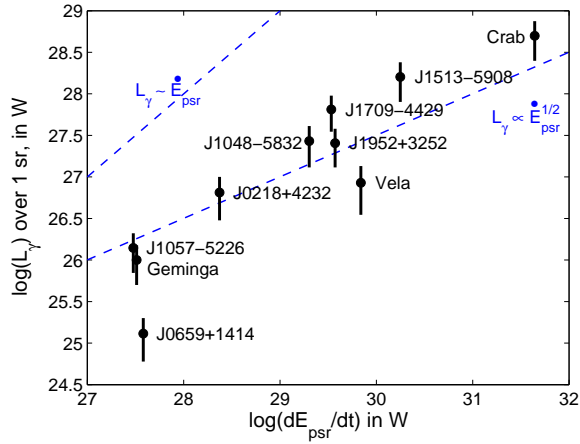}
\caption{High-energy luminosity (as measured over 1 sr above 1 keV)
vs. spindown power as in \cite{thompson04}.}} \label{figlum}
\end{figure}

\section{Clues from observations}

\begin{figure}[t]
{\includegraphics[width=1.0\textwidth]{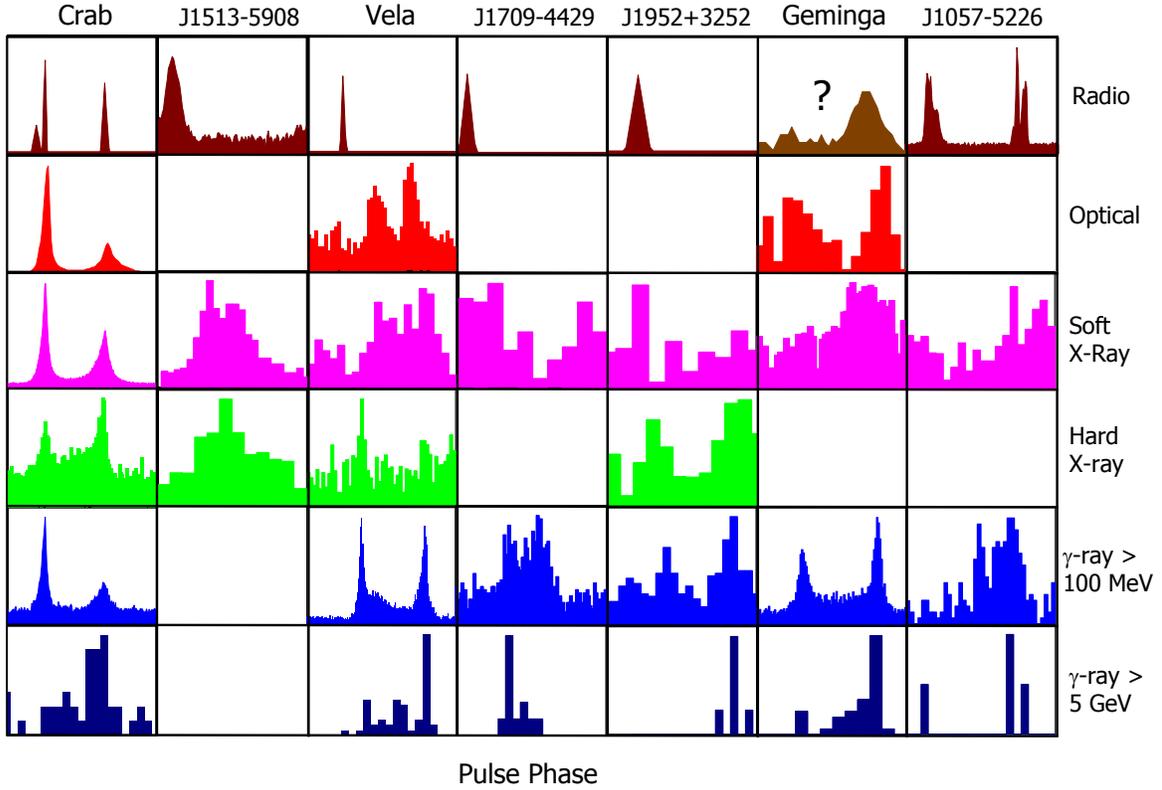}
\caption{Light curves of seven $\gamma$-ray pulsars in six energy bands, 
as in Thompson \cite{thompson04}.}} \label{figltc}
\end{figure}

Comparing the lightcurves at different wavelengths for a single
object potentially provides a wealth of information on the gap
location and on the cascade development and its radiation
processes. Spectral cut-offs in $\gamma$ rays also bear signatures
of the pair production mechanism. Compton up-scattered radiation
at TeV energies can constrain the number of secondary pairs and
their location. The overall luminosity is linked to the rotation
power and the maximum $E_{\parallel}$ field that is not screened
along B. Unfortunately, only 9 pulsars (including 1 ms pulsar) are
known in $\gamma$ rays, 10 in the optical and 30 in X rays, and
they present an outstanding variety of lightcurve configurations 
and spectral shapes (see Figure \ref{figltc}). Crab-like objects
and several ms pulsars exhibit well synchronized pulses over many
decades in energy that suggest very short cascades (within
hundreds of meters) or an origin in the caustic zone, whereas the
complex and out-of-phase peaks and bumps of Vela-like or PSR
B1055-52-like objects suggest the presence of several beams or of
a heterogeneous beam with strongly varying spectra. The detection
in $\gamma$ rays of 6 of the 9 radio pulsars with highest
$\dot E_{\rm psr}$-over-square-distance rank indicates a close relationship
between the onset of high-energy showers and coherent radio
emission, but the existence of Geminga, the only radio-quiet
pulsar known so far, proves that the radio and high-energy beams
have different apertures or directions. Only the younger ($< 0.4$
Myr), brighter pulsars have been detected at high energy so far,
yet they illustrate how important it is to understand the
acceleration and cascading processes since the $\gamma$-ray
luminosity clearly dominates the radiation budget. It exceeds the
radio luminosity by 6 or 7 orders of magnitude.

$\gamma$-ray pulsars behave somewhat counter-intuitively: the
older pulsars have harder spectra and are more efficient in
$\gamma$ rays. Figure \ref{figlum} shows that the $\gamma$-ray
luminosity over 1 sr scales as $L_{\gamma 1sr} \propto
\dot E_{\rm psr}^{1/2}$ over 4 decades in $\dot E_{\rm psr}$ (\cite{thompson04}),
despite the likely dispersion in the true beam apertures. This
trend suggests that the cascade current is a constant fraction of
the maximum GJ current across the polar caps ($\dot{N}_{GJ} =
\rho_{GJ} \pi R^2 \theta_{PC}^2/e \simeq 1.4 \times 10^{32}
e^{\pm} s^{-1} (P/0.1 s)^{-2} (B_p/10^8 T)$) and that the maximum
energy $E_{max}$ gained in the gap is rather constant and radiated
away by the cascade. As discussed above, the feedback between
particle acceleration and electrical screening by the cascading
yields rather stable $E_{max}$ around 10 TeV in both the polar cap
\cite{harding98} and outer gap \cite{hhs03} models. The relation
should break for $\dot E_{\rm psr} < 10^{26}$ W for the
$L_{\gamma}/\dot E_{\rm psr}$ efficiency not to exceed 100\%. For older
pulsars, the relation should turn to $L_{\gamma 1sr} \propto
\dot E_{\rm psr}$ for both polar cap and outer gap accelerators because of
the inefficient electrical screening and of the gap filling a
large fraction of the open magnetosphere. Cascading being less efficient,
one expects emission from older pulsars to be dominated by hard
curvature radiation. This is the case for instance in ms pulsars
where, because of the low magnetic field, the unscreened
$E_{\parallel}$ keeps accelerating particles to high altitudes and
the resulting curvature radiation, radiation-reaction limited to
several tens of GeV, could provide a measure of $E_{\parallel}$.

Spectral signatures at very high energies will help distinguishing
between models. The overall cascade spectrum and potential
asymmetries in the cut-off energy between the leading and trailing
side are not discriminant, but the sharpness of the cut-off and
its dependence on the magnetic field can be. Emission from the
polar cap should abruptly break at energies $E_{cut} \propto
P^{1/2} B_p^{-1/2} (r/R)^{7/2}$ \cite{hardingheidel05}. The
observed dependence between $E_{cut}$, from 20 MeV for PSR
B1509-58 to more than 10 GeV for PSR J1951+32, and the surface
magnetic field $B_P$ \cite{thompson04} is compatible with a magnetic origin of the
cascade, but the dispersion is large and any conclusion requires
more data. Observations above 100 GeV with the HESS telescope
cannot distinguish between an exponential and a super-exponential
break, yet they start to constrain the amount of inverse Compton
radiation produced in the thick outer gap, for instance in Vela
\cite{schmidt04}.

Population studies offer a statistical means to test the radio and
$\gamma$-ray luminosity dependence on pulsar age and power, and of
the aperture and sweeping properties of the beams. It is evident
that a pencil beam from the polar cap, a funnel beam from the slot
gap, and a fan beam from the outer gap will sweep differently
across the sky and yield different numbers of radio-quiet and
radio-loud objects if the radio beam significantly differs from
the $\gamma$-ray one. This test, however, may turn out to be less
conclusive for young pulsars which have high-altitude radio conal
beams closer to the high-energy beams \cite{manchester96,jw06}.
The current studies of radio-quiet and radio-loud statistics using
the polar cap and the slot gap models are compatible with the
detection by EGRET of 8 radio pulsars, the existence of Geminga,
the presence of a score of unidentified EGRET sources along
the Galactic plane and of a handful of sources in the Gould Belt
\cite{gonthier04,gonthier05}. The outer gap can also contribute a
large fraction of the unidentified sources at low latitudes
\cite{zhang05}. For all models, only the younger (brighter)
$\gamma$-ray pulsars can be detected above the intense emission
from the Milky Way.

In conclusion, many fundamental questions remain open in pulsar
twinkling. Observing and modelling a large sample of them at
$\gamma$-ray and hard X-ray energies with the forthcoming
telescopes (AGILE, GLAST, SIMBOL-X) and trying to get polarization
data at high energy (PoGO) will bring critical clues and will
nicely complement the studies of binary pulsars to understand the
electrodynamics of large magnetic fields in strong gravity fields.

\begin{theacknowledgments}
We warmly thank Dave Thompson and Jarek Dyks for their help and time.
\end{theacknowledgments}

\end{document}